\documentclass[journal]{IEEEtran}
\usepackage{newtxtext}
\usepackage{newtxmath}
\usepackage{amsmath,amsfonts}
\usepackage{algorithmic}
\usepackage{algorithm}
\usepackage{array}
\usepackage[caption=false,font=normalsize,labelfont=sf,textfont=sf]{subfig}
\usepackage{textcomp}
\usepackage{stfloats}
\usepackage{url}
\usepackage{verbatim}
\usepackage{graphicx}
\usepackage{cite}
\usepackage[section]{placeins}
\usepackage[switch]{lineno}
\usepackage{flafter}  
\usepackage[T1]{fontenc}
\usepackage{hyperref}
\usepackage{multirow, multicol, booktabs}

\hypersetup{%
  colorlinks=true,
  linkcolor=blue,
  citecolor=teal,
  filecolor=magenta,
  urlcolor=teal,
  pdftitle={Al-STJs for Nucl. Spectroscopy},
  pdfpagemode=FullScreen,
}

\graphicspath{{../out/5DAT-2022-06-10-32Al70-7Be/}{../in/figures/}{../out/}} 
\begin{document}

\title{Aluminum-Based Superconducting Tunnel Junction Sensors for Nuclear Recoil Spectroscopy}

\author{%
  Spencer~L.~Fretwell\IEEEauthorrefmark{1}\IEEEauthorrefmark{2},%
  \ Connor~Bray\IEEEauthorrefmark{1},%
  \ Inwook~Kim\IEEEauthorrefmark{2},%
  \ Andrew~Marino\IEEEauthorrefmark{1},%
  \ Benjamin~Waters\IEEEauthorrefmark{1}\IEEEauthorrefmark{14},%
  \ Robin~Cantor\IEEEauthorrefmark{3},%
  \ Ad~Hall\IEEEauthorrefmark{3},%
  \ Pedro~Amaro\IEEEauthorrefmark{4},%
  \ Adrien~Andoche\IEEEauthorrefmark{5},%
  \ David~Diercks\IEEEauthorrefmark{6},%
  \ Abigail~Gillespie\IEEEauthorrefmark{1},%
  \ Mauro~Guerra\IEEEauthorrefmark{4},%
  \ Cameron~N.~Harris\IEEEauthorrefmark{1}\IEEEauthorrefmark{13},%
  \ Jackson~T.~Harris\IEEEauthorrefmark{7},%
  \ Leendert~M.~Hayen\IEEEauthorrefmark{8},%
  \ Paul~Antoine~Hervieux\IEEEauthorrefmark{9},%
  \ Geon~Bo~Kim\IEEEauthorrefmark{2},%
  \ Annika~Lennarz\IEEEauthorrefmark{10}\IEEEauthorrefmark{15},%
  \ Vincenzo~Lordi\IEEEauthorrefmark{2},%
  \ Jorge~Machado\IEEEauthorrefmark{4},%
  \ Peter~Machule\IEEEauthorrefmark{10},%
  \ David~McKeen\IEEEauthorrefmark{10},%
  \ Xavier~Mougeot\IEEEauthorrefmark{5},%
  \ Francisco~Ponce\IEEEauthorrefmark{11},%
  \ Chris~Ruiz\IEEEauthorrefmark{10},%
  \ Amit~Samanta\IEEEauthorrefmark{2},%
  \ José~Paulo~Santos\IEEEauthorrefmark{4},%
  \ Joseph~Smolsky\IEEEauthorrefmark{1}\IEEEauthorrefmark{3},%
  \ Caitlyn~Stone-Whitehead\IEEEauthorrefmark{1},%
  \ Joseph~Templet\IEEEauthorrefmark{1},%
  \ Wouter~Van~De~Pontseele\IEEEauthorrefmark{1},%
  \ William~K.~Warburton\IEEEauthorrefmark{7},%
  \ K.~G.~Leach\IEEEauthorrefmark{1}\IEEEauthorrefmark{12},%
  \ S.~Friedrich\IEEEauthorrefmark{2}%
  \thanks{\IEEEauthorrefmark{1}Department of Physics, Colorado School of Mines;~1523 Illinois St, Golden, 80401, Colorado, USA}%
  \thanks{\IEEEauthorrefmark{2} (E-Mail: \href{mailto:sfretwel@mines.edu}{sfretwel@mines.edu})}
  \thanks{\IEEEauthorrefmark{2}Lawrence Livermore National Laboratory;~7000 East Ave, Livermore, CA 94550, USA}%
  \thanks{\IEEEauthorrefmark{4}LIBPhys, LA-REAL, NOVA FCT, Universidade NOVA de Lisboa;~2829-516 Caparica, Portugal}
  \thanks{\IEEEauthorrefmark{3}STAR Cryoelectonics LLC;~25A Bisbee Ct, Santa Fe, NM 87508}%
  \thanks{\IEEEauthorrefmark{5}Université Paris-Saclay, CEA, List, Laboratoire National Henri Becquerel (LNE-LNHB);~91120 Palaiseau, France}%
  \thanks{\IEEEauthorrefmark{6}Shared Instrumentation Facility, Colorado School of Mines;~1523 Illinois St, Golden, 80401, Colorado, USA}%
  \thanks{\IEEEauthorrefmark{7}XIA LLC;~2744 E 11th St, Oakland, CA 94601}%
  \thanks{\IEEEauthorrefmark{8}LPC Caen, ENSICAEN, Université de Caen, CNRS/IN2P3;~14000 Caen, France}%
  \thanks{\IEEEauthorrefmark{9}Université de Strasbourg, CNRS, Institut de Physique et Chimie des Matériaux, 23 Rue du Loess Bâtiment 69, 67200 Strasbourg, France}%
  \thanks{\IEEEauthorrefmark{10}TRIUMF;~4004 Wesbrook Mall, Vancouver, BC V6T 2A3, Canada}%
  \thanks{\IEEEauthorrefmark{11}Pacific Northwest National Laboratory;~Richland, WA 99354, USA}%
  \thanks{\IEEEauthorrefmark{12}Facility for Rare Isotope Beams, Michigan State University;~640 S Shaw Lane, East Lansing, 48824, Michigan, USA}%
  \thanks{\IEEEauthorrefmark{13}Ciambrone Radiochemistry Lab;~Patrick Space Force Base, FL 32925}%
  \thanks{\IEEEauthorrefmark{14}Maybell Quantum;~7100 Broadway, Building 3, Denver, CO 80221}%
  \thanks{\IEEEauthorrefmark{15}Department of Physics and Astronomy, McMaster University;~Hamilton, Ontario L8S 4M1, Canada}%
  \thanks{Manuscript submitted September 26, 2025}
} 



\maketitle

\begin{abstract}
  The BeEST experiment is searching for sub-MeV sterile neutrinos by measuring nuclear recoil energies from the decay of \( ^7 \)Be implanted into superconducting tunnel junction (STJ) sensors.
  The recoil spectra are affected by interactions between the radioactive implants and the sensor materials.
  We are therefore developing aluminum-based STJs (Al-STJs) as an alternative to existing tantalum devices (Ta-STJs) to investigate how to separate material effects in the recoil spectrum from potential signatures of physics beyond the Standard Model.
  Three iterations of Al-STJs were fabricated.
  The first had electrode thicknesses similar to existing Ta-STJs.
  They had low responsivity and reduced resolution, but were used successfully to measure \( ^7 \)Be nuclear recoil spectra.
  The second iteration had STJs suspended on thin SiN membranes by backside etching.
  These devices had low leakage current, but also low yield.
  The final iteration was not backside etched, and the Al-STJs had thinner electrodes and thinner tunnel barriers to increase signal amplitudes.
  These devices achieved 2.96~eV FWHM energy resolution at 50 eV using a pulsed 355~nm (\( \sim \)3.5 eV) laser.
  These results establish Al-STJs as viable detectors for systematic material studies in the BeEST experiment.
\end{abstract}

\begin{IEEEkeywords}
  Superconducting tunnel junctions, nuclear recoil spectroscopy, sterile neutrinos, beryllium-7, aluminum STJs, superconducting device radiation effects, nanofabrication, laser calibration, BeEST experiment
  \vspace{-2em}
\end{IEEEkeywords}


\section{Introduction}%
\label{intro}

\IEEEPARstart{P}{hysics} in the neutrino sector offers opportunities for new discoveries Beyond the Standard Model (BSM) of particle physics.
The origins of neutrino mass and the complete picture of neutrino interactions remain unknown, leaving open fundamental questions about the existence of right-handed sterile neutrinos~\cite{DasguKopp2021-PR}, the mass of active neutrinos~\cite{Xing2020-PR}, and the nature of neutrino wavepackets~\cite{BeEST-SmolsLeach2025-N}.

The Beryllium Electron capture in Superconducting Tunnel junctions (BeEST) experiment investigates fundamental neutrino properties through their correlations with recoil energies from nuclear decays~\cite{LeachFried2022-JLTP}.
Specifically, the BeEST uses superconducting tunnel junctions (STJs), which provide eV-scale energy resolution and high speed for high-statistics measurements, to study the electron capture (EC) decay of implanted \( ^7 \)Be and directly correlate the daughter recoil energy with the neutrino mass.
The spectra are calibrated with a pulsed UV laser with a precision \(<\)~0.01 eV so that the experiment can resolve eV-scale shifts in nuclear recoils that correspond to hypothetical sub-MeV neutrino masses predicted by particular Seesaw mechanisms~\cite{XingZhou2011-ZUP, BoyarDrewe2019-PPNP}.
This framework can also constrain the spatial width of the neutrino wave-packet, which may contribute to the observed broadening of the recoil peaks~\cite{BeEST-SmolsLeach2025-N}.

The BeEST Collaboration has demonstrated this technique successfully with tantalum-based STJs (Ta-STJs) that were originally designed for soft X-ray spectroscopy~\cite{CarpeFried2013-ITAS}.
The initial campaign produced the leading laboratory limits on 100--850~keV right-handed neutrinos~\cite{FriedKim2021-PRL}, with the current blinded analysis promising order-of-magnitude improvements~\cite{BeEST-KimBray2025-PRD}.
However, systematic uncertainties from detector material properties and \( ^7 \)Be implantation site effects affect the experimental sensitivity.
To address these limitations, the collaboration is investigating material effects through simulations~\cite{SamanFried2023-PRA}, destructive assays of \( ^7 \)Be and \( ^7 \)Li dopant distributions~\cite{Harris2024-CSoM}, and development of alternative STJ absorber materials (this work).

\IEEEpubidadjcol%

Alternative-material STJs would provide a complementary testbed to investigate features in the spectral lineshape that are a result of materials effects rather than new physics.
In particular, the BeEST spectrum includes several features which originate from the atomic configuration of \( ^7 \)Be, its  \( ^7 \)Li daughter, and their interactions with the superconducting sensor materials.
For example, the centroids, linewidths, and peak shapes for these features differ from the values expected for the decay of isolated \( ^7 \)Be atoms~\cite{BeEST-KimBray2025-PRD}.
It is therefore desirable to repeat the BeEST experiment with STJs made from different materials to systematically test for subtle material-dependent effects.

\section{Aluminum STJs}%
\label{al}


The initial motivation for aluminum-based STJs (Al-STJs) had been their potential for higher energy resolution, because that would allow for the detection of smaller peak shifts associated with lower sterile neutrino masses. Aluminum's smaller superconducting band gap (\( \Delta_\text{Al} \approx 0.18 \)~meV) increases the number of signal charges per unit energy compared to tantalum (\( \Delta_\text{Ta} \approx 0.68 \)~meV) by a factor of \( \sim \)4 and therefore improves the limiting energy resolution by a factor of \( \sim \)2.
This turns out to be less relevant for the BeEST experiment because the signal peaks are broadened beyond the energy resolution of the STJs~\cite{FriedKim2021-PRL, BeEST-KimBray2025-PRD, SamanFried2023-PRA}. It may, however, be relevant for future experiments where higher energy resolution is desirable.

The current motivation for Al-STJs for the BeEST experiment is to study the effects of the host material on the details of the decay spectra.
The interaction of implanted \( ^7 \)Be with the STJ absorber material likely affects the centroid energies and the peak structures in the recoil measurement.
For example, the tailing and energy positions of peaks due to electronic processes such as Auger emission, shake-up, and shake-off will be different in Al because of differences in hybridization.
Further, the Doppler broadening of peaks from the decay of \( ^7 \)Be through a short-lived nuclear excited state, \( ^7 \)Li\( ^\star \), may be different in Al compared to Ta.
Since the magnitude of this broadening is related to the energy attenuated by the second recoil following the decay of \( ^7 \)Li\( ^\star \), the peak width is a measure of the slowdown timescale relative to this state's lifetime.
Aluminum’s closer mass to \( ^7 \)Li\( ^\star \) and closer packing compared to tantalum may reduce this timescale, leading to narrower peaks.
This in particular is exciting for the BeEST heavy neutrino search, as the spaces between peaks are the most sensitive regions for this search~\cite{FriedKim2021-PRL}, but they are currently dominated by counts within the nuclear excited state (ES) peaks.
In summary, a comparative study of \( ^7 \)Be decay spectra in Al- and Ta-STJs provides a systematic test of material-dependent effects.

One difficulty for Al-based STJs is that is that the small Al gap prevented the use of a quasiparticle trap in this work.
In Ta-STJs, an Al tri-layer is used to trap quasiparticles by inelastic scattering, confining them near the aluminum oxide tunnel barrier and thereby enhancing the tunneling rate.
The Ta absorbers are therefore thick for high energy absorption efficiency, while the Al traps are thin for fast tunneling.
In contrast, quasiparticles in the Al-STJs are allowed to diffuse throughout the entire Al-STJ absorber,
which reduces their tunneling probability and offsets the factor \( \sim \)4 gain in quasiparticle production in Al compared to Ta.
For Al-STJs, fabrication of a lower-gap trap region would have required the integration of smaller-gap superconducting materials such as hafnium into the existing Al-based process flow, requiring substantial development effort.
Further, low-noise operation of such devices would have required lower base temperatures than are capable using our adiabatic demagnetization refrigerator (ADR) systems.
The choice of Al electrode thickness was therefore a careful trade-off between energy absorption efficiency at the energy range of interest and tunneling rate.



\subsection{Design}%
\label{al.design}

The design of Al-STJs is informed by the design of earlier Ta-STJs~\cite{CarpeFried2013-ITAS}. STJ geometries and areas, array sizes, and chip sizes are similar to these Ta-STJs, with additional features designed in consideration of the energy range of interest and secondary radiations produced by \( ^7 \)Be.
Three STJ sizes of (70~\( \mu \)m)\(^2\), (130~\( \mu \)m)\(^2\), and (200~\( \mu \)m)\(^2\) balance the competing needs for large active area and high energy resolution.
The pixels are rotated by 45\( ^\circ \) relative to the chip edges to improve the suppression of the DC Josephson current and allow operating multiple pixels simultaneously despite small differences of suppression~\cite{Peterson1991-C, HouwmGijsb1991-PCS}.
Since the Al-STJs do not have a trapping layer, a larger-gap niobium (\( \Delta_\text{Nb} \approx \) 1.5~meV) plug is used in the ground traces to confine thermalized quasiparticles at the aluminum gap energy, preventing crosstalk between pixels through the shared ground.

The chips were designed to be backside etched to remove substrate below the pixels and suspend them on thin membranes.
Each pixel is placed on a separate membrane that exceeds the STJ dimensions by 10~\( \mu \)m on all sides. This membrane suppresses the spectral background from substrate phonons, in particular those produced by the \( \gamma \)-rays emitted by \( ^7 \)Li\(^\star \).
To ensure structural stability of the remaining material, the substrate grid between the membranes is at least 50~\( \mu \)m wide. Octagonal SU-8 posts can also be deposited to allow for mounting Si collimators directly onto the chip for use during \( ^7 \)Be implantation or laser calibration.

\begin{figure}[b]
  \centering
  \vspace{-0.5em}
  \includegraphics[width=1\linewidth]{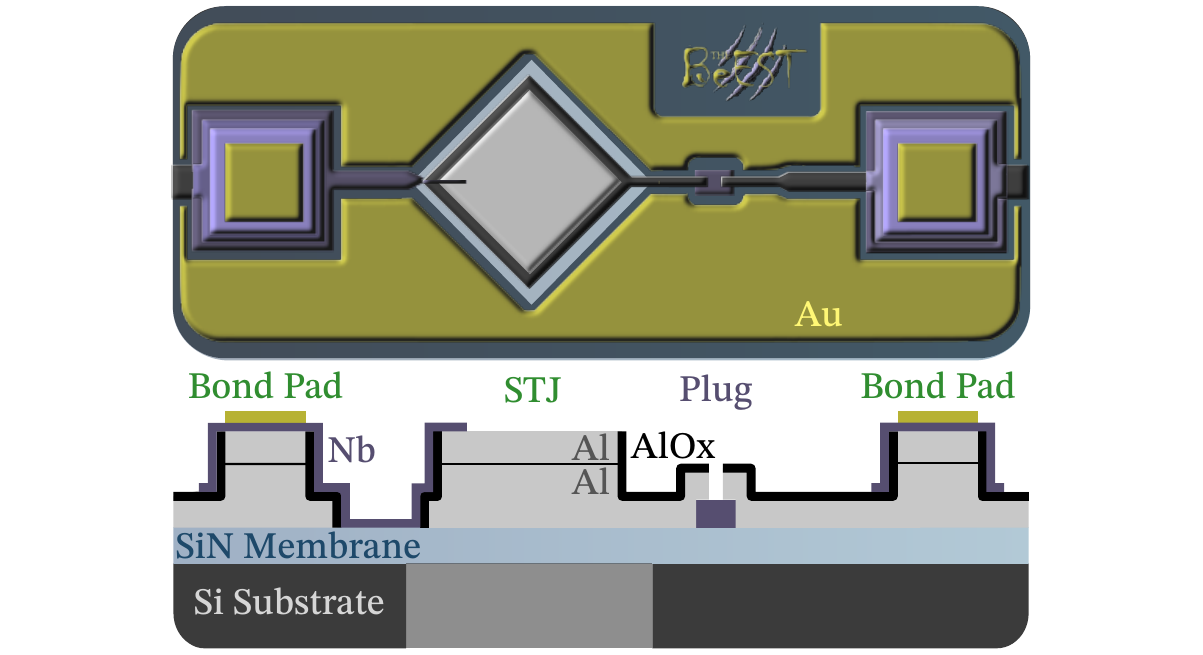}%
  \vspace{-0.5em}
  \caption{%
    Top-view and cross-sectional primitive of an Al-AlOx-Al STJ on an SiN membrane with Nb wiring, Nb quasiparticle plugs, Au bond pads, and Au thermalization layer.
    This primitive can be duplicated to produce chips with arrays of STJ pixels.
    The Al base electrode is read out by the right pad and requires a Nb plug.
    Anodization of the junction edges (see Section~\ref{al.fabrication}) is shown in black, and is used to insulate the wiring from the base electrode when reading out the top electrode using the left pad.
    Layers not to scale.%
    \label{fig:design-schematic}
  }%
\end{figure}

A top-down render ``primitive'' with  minimum features of an Al-STJ---electrodes, wiring, plug, and bond pads---is shown in Figure~\ref{fig:design-schematic}.
This primitive was duplicated to form small arrays with 32 STJ pixels, to match the 32-channel MPX-32D preamplifiers from XIA LLC~\cite{WarbuHarri2015-NIMA} that were used in this work to read out and control the devices.
Larger 128-pixel arrays, which are roughly (1 cm)\(^2\) in size, were also designed to maximize the capability of the XIA readout system, which can hold four preamplifier cards per module.
Groups of eight pixels share a single ground wire to reduce the number of readout wires to the cryostat cold stage.
Finally, 14-pixel chips were designed as test devices to compare the performance of different pixel sizes and examine the impact of the SiN membranes.

\subsection{Fabrication}%
\label{al.fabrication}%

The Al-STJs are fabricated by STAR Cryoelectronics LLC on 4'' wafers using a proprietary selective aluminum etch process.
The p-doped Si \( \langle \)100\( \rangle \) wafers with resistances on the order of 10~\( \Omega \)~cm are purchased with a 500~nm low-stress silicon nitride (SiN) layer from Addison Engineering.
The Nb plugs are deposited first by lift-off to ensure good electrical contact with the Al wiring.
Next, the Al tri-layer is deposited in a single vacuum cycle, with the oxidation conditions chosen for a critical current density \( j_C = 50 \)~A/cm\( ^{2} \).
The STJ geometries are defined by ion milling in two stages.
The first milling step defines the Al base layer and ground traces, and the second step defines the junction area.
The tri-layer is also retained at the bond pads to increase their thickness and improve the adhesion of bond wires.
Before the photoresist is removed, the junction edges are anodized to passivate the STJ side walls and suppress leakage currents.
Then, the resist is stripped and the Nb wiring to the top electrode is deposited and patterned by lift-off.
Finally, a gold film is deposited onto the bond pads to provide a low-resistance interface for wire bonding.
Au is also deposited on most surfaces 10~\( \mu \)m away from other features to act as a sink for athermal phonons in the Si substrate and thereby assist in the suppression of the \( \gamma \)-induced background.
If desired, the wafer may then be etched from the backside to suspend the pixels on SiN membranes.
Figure~\ref{fig:design-schematic} (bottom) shows a cross section of the layers---not to scale---that result from this process. Each fabrication run produces 42 128-pixel chips, 130 32-pixel chips, and 98 test chips.



Three Al-AlOx-Al STJ wafers were fabricated and tested during the development of the devices.
The parameters which changed between each fabrication and the results of experimental testing are summarized in Table~\ref{tab:fab_comparison}.


\section{Testing and Results}%
\label{exp}

The Al-STJs were tested in a ``wet'' two-stage ADR at Lawrence Livermore National Laboratory pre-cooled by liquid N\(_\text{2}\) and liquid He, with a primary (guard) stage cooled by gadolinium gallium garnet (\( T_\text{base, GGG}\approx1\)~K) and a secondary (detector) stage cooled by paramagnetic ``ferric ammonium alum'' (FAA).
They were operated at a temperature of \( \sim \)100~mK in a magnetic field of \( \sim \)100 Gauss in the direction of the junction diagonal to suppress the DC Josephson current and the Fiske modes.
The current-voltage characteristics were measured with a transimpedance amplifier from XIA LLC at room-temperature (e.g. Fig.~\ref{fig:iv-fab1})~\cite{WarbuHarri2015-NIMA}.


The STJ response was tested with a pulsed 355~nm laser (Spectra-Physics \verb`J40-BL6-106Q`) as an energy calibration source. The laser was attenuated to deliver 1--30 photons per pulse per STJ, which produced a series of photopeaks in each STJ spaced at integer multiples of the single photon energy of \(3.49865 \pm 0.00015\)~eV~\cite{PonceSwanb2018-PRC}.
The responsivity is then measured in units of nA of signal current per keV of absorbed energy.
The noise is set by the shot noise of the bias current \( I_\text{bias} \) at \( V_\text{bias} \), the Johnson  noise of the 1~M\( \Omega \) feedback resistor \( R_\text{F} \) at room temperature, and the noise contribution of the preamplifier. The BF862 JFET at the preamplifier input has a voltage noise \( e_\text{n, FET} \) of 0.8~nV/\(\sqrt{\text{Hz}}\)~\cite{WarbuHarri2015-NIMA}, which is scaled by the dynamic resistance, \(R_\text{dyn}\), of the STJ.\@
These noise sources add in quadrature to produce a total current noise of:
\begin{equation}
  i_\text{n, total}^2 = 2 e {I_\text{bias}}  + {\left(\frac{e_\text{n, FET}}{R_\text{dyn}}\right)}^2 + \frac{4 k_\text{B} T}{R_\text{F}}.
\end{equation}

Selected STJ chips were implanted with \( ^7 \)Be through a Si collimator at the Isotope Separator and Accelerator facility at TRIUMF in Vancouver, Canada~\cite{DilliKruck2014-HI}.
Proton bombardment of a uranium carbide target produced a wide range of radioactive nuclei via spallation reactions.
\(^7\)Be was selectively ionized using resonant laser ionization, mass-selected, accelerated, and delivered to the Ion Implantation Station. Following implantation, the devices were rinsed to remove residual surface activity, and the implanted activity of \( ^7 \)Be was measured via the intensity of the 478~keV \(\gamma \)-ray.

\subsection{First Wafer Testing}%
\label{exp.fab1}


The first Al-AlOx-Al STJs were fabricated on SiN-coated wafers with 265~nm thick base electrodes and 200~nm thick top electrodes.
The electrode thicknesses match those of earlier Ta-STJs and were chosen to test compatibility with the existing process flow. A second goal was to test the growth of aluminum on SiN. No backside etching was performed.

\begin{figure}[!bp]
  \vspace{-1em}
  \includegraphics[width=\linewidth]{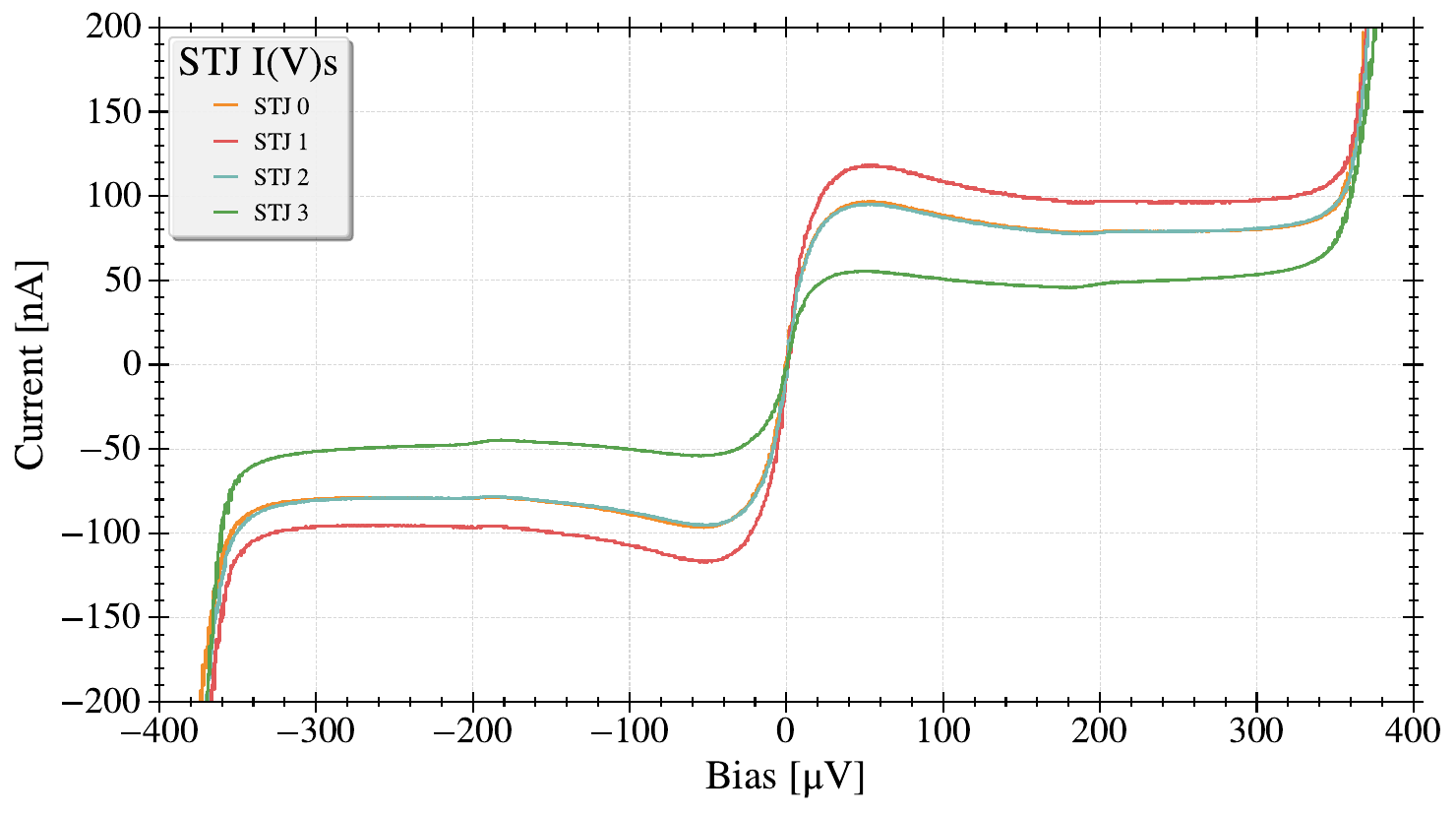}
  \caption{%
    The current-voltage characteristics of the first Al-STJs on SiN at 100~mK have very high dynamic resistance above \( 180~\mu \)V and surprisingly high thermal currents of 40--100~nA.%
    \label{fig:iv-fab1}
  }
\end{figure}

\( I(V) \) characteristics from these first Al-STJs show a sharp increase at a voltage of \(360~\mu \)V, which is close to the nominal gap \(2\Delta_\text{Al}/e \) for bulk Al.
No significant increase in leakage current at \( 1\Delta_\text{Al}/e \) is observed, indicating high junction quality and limited flux trapping during this test.
Despite the operating temperature of \( \sim \)100~mK, these STJs showed a surprisingly high thermal contribution to the leakage current, but also a rather high dynamic resistance between 180~\( \mu \)V and 300~\( \mu \)V.

\begin{figure}[!tbp]
  \centering
  \includegraphics[width=\linewidth]{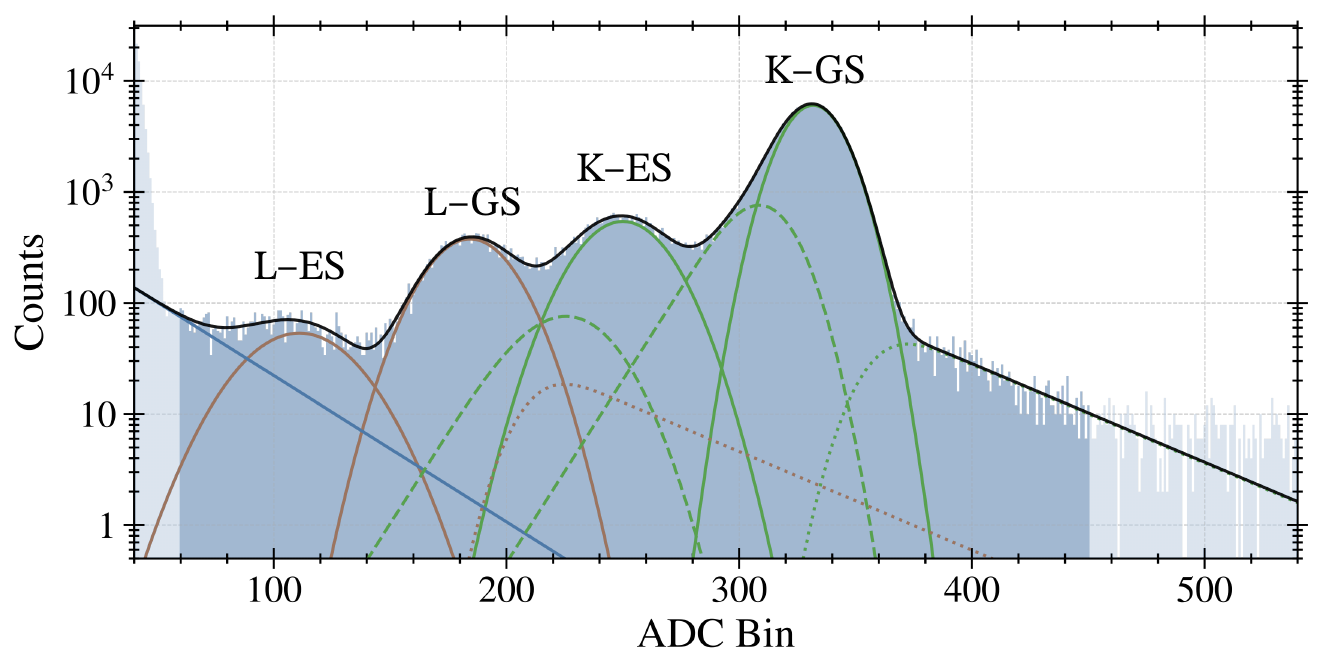}%
  \vspace{-1em}
  \caption{%
    First-run Al-STJ nuclear spectrum and fit to the BeEST model.
    The devices demonstrate \( \sim \)8~eV of electronic noise, but nonetheless successfully resolve the primary peaks expected from \( ^7 \)Be EC decay (labeled, see text for description).%
    \label{fig:gen1-nuclear-spectrum}
  }
  \vspace{-1em}
\end{figure}

The STJs with the \( I(V) \)s shown in Figure~\ref{fig:iv-fab1} were biased at a voltage of 180~\( \mu \)V where the dynamic resistance of the STJ is close to infinite to suppress the JFET noise. Under these conditions, shot noise contributes \( \sim \)0.16~pA/\(\sqrt{\text{Hz}}\) and Johnson noise \( \sim \)0.13~pA/\(\sqrt{\text{Hz}}\).
In quadrature, the terms yield a total current noise of \( \sim \)0.22~pA/\(\sqrt{\text{Hz}}\). The measured baseline noise of \( \sim 0.26\)~pA/\(\sqrt{\text{Hz}}\) is somewhat higher for reasons that are currently being investigated.
Still, the high junction quality confirms the feasibility to fabricate Al-STJs on SiN.

\subsection{Nuclear Recoil Spectroscopy Results}%
\label{exp.fab1nuc}

The devices were implanted at TRIUMF at a beam energy of 10~kV with \( \sim \)1~Bq per pixel of \(^7\)Be with a \( \sim \)13:1 ratio of lithium to beryllium.
\(^7\)Li, which is surface-ionized in the hot target environment, may be suppressed by operating TRIUMF's Ion Guide Laser Ion Source (IG-LIS) in suppression mode, however for this implantation, IG-LIS was limited to transmission mode operation.
The STJs exhibited no degradation in electrical characteristics after implantation, with similarly high dynamic resistances and elevated thermal currents.
The preamplifier output in response to the decay of \( ^7 \)Be was Gaussian filtered with an analog spectroscopy amplifier (Ortec \verb`672`) with a 10~\( \mu \)s shaping time and captured by a nuclear MCA (Ortec \verb`927 ASPEC`).
Spectra were acquired for multiple ADR cycles over the course of a week.

Pulse amplitudes were quite small due to the relatively thick electrodes.
The responsivity was approximately \( \sim \)20 nA/keV, roughly a factor 8 smaller than the Ta devices~\cite{BeEST-KimBray2025-PRD}.
This caused the electronic noise of \( \sim \)8 eV to dominate the energy resolution, and as a result, these first Al-STJs could not resolve the 3.5~eV-spaced peaks of the calibration laser.

An uncalibrated sum spectrum from five runs is shown in Figure~\ref{fig:gen1-nuclear-spectrum}.
It shows the four primary peaks from the EC decay of \( ^7 \)Be:
the K-GS peak for EC of the 1\(s \) (K-shell) electron and decay into the nuclear ground state (GS) of \( ^7 \)Li;
the K-ES peak for K-capture decay into the excited nuclear state (ES) of \(^7\)Li\(^\star \);
and the L-GS and L-ES peaks for the corresponding 2\(s \) (L-shell) capture processes.
The hole relaxation after K-capture, which emits an Auger electron, occurs on timescales much shorter than the signal rise time so that the K-capture peaks are \( \sim \)50 eV higher in energy than the L-capture peaks.

The widths of these peaks, presented as a percent ratio of their ADC width to their ADC position, are 8.5\% for the K-GS peak, 16\% for the K-ES peak, and 21\% for the L-GS peak.
An approximate calibration may be done assuming the centroid values are similar to those measured in Ta-STJs with high accuracy~\cite{BeEST-KimBray2025-PRD}.
This calibration produces widths of \( \sim \)9.8~eV FWHM for the K-GS peak at 108.5 eV, \( \sim \)14.2 eV FWHM for the K-ES peak at 81.3 eV, and \( \sim \)13.5 eV FWHM for the L-GS peak at 56.9 eV.
These numbers show that the L-GS peak is wider than the K-GS peak in Al-STJs, which also occurs in Ta-STJs and is currently not understood~\cite{BeEST-KimBray2025-PRD}.
They also show Doppler broadening of the ES peaks (Section~\ref{al}).
In addition, tails are visible below the K-peaks, which arise from Auger electron escape.
The tail above the K-GS peak results primarily from L-electron shake-off after K-capture.
At low energies, background from Be implantation between pixels is visible.

These widths also provide the first hint that interactions between the implants and the sensor materials do affect spectral details, since the Doppler-broadened K-ES peak is narrower in Al-STJs than in Ta-STJs, although the resolution in Al-STJs is much worse. The reduced Doppler broadening likely reflects a faster thermalization of the recoiling \( ^7 \)Li nucleus in Al-STJs compared to Ta-STJs. Devices with higher energy resolution are required to quantify systematic differences between Al and Ta-STJs.

\begin{figure}[!bt]
  \centering
  \includegraphics[width=.7\linewidth]{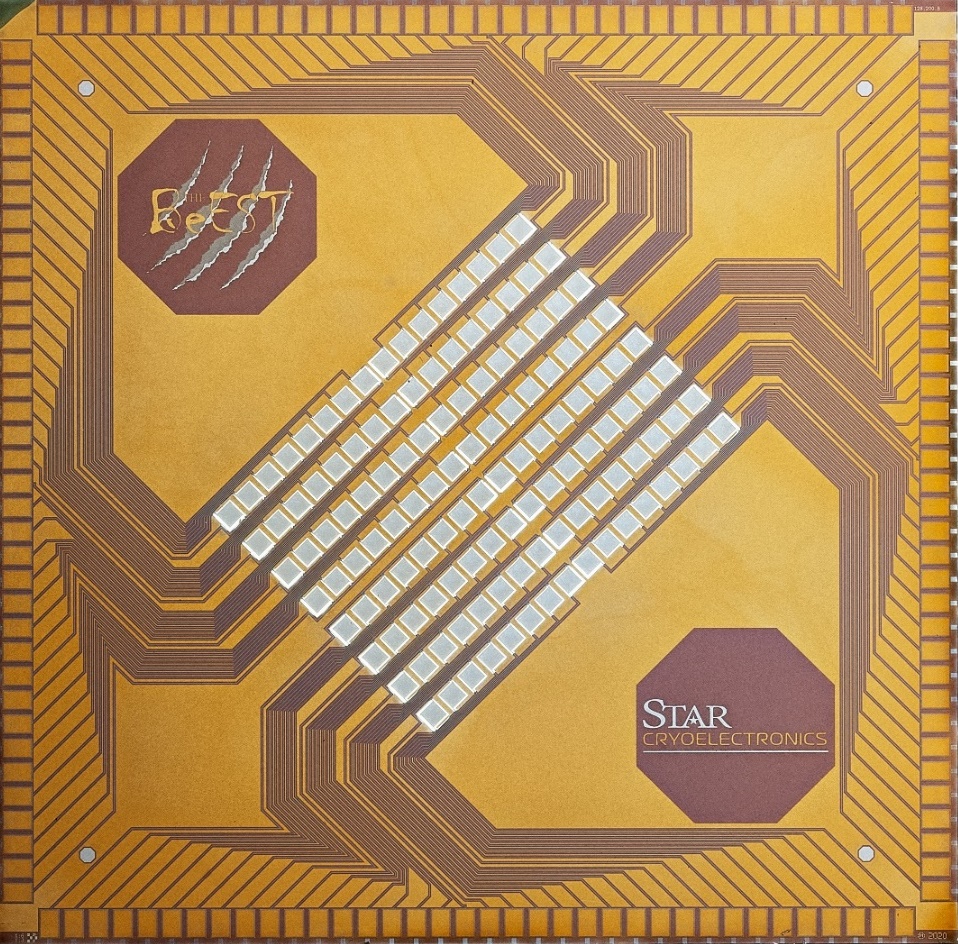}
  \caption{%
    128-pixel array of (130~\( \mu \)m)\(^2\) STJ sensors on SiN membranes.
    The array is back-lit to highlight the transparent membranes.%
    \label{fig:fabricated-chip}
  }
  \vspace{-1em}
\end{figure}

\subsection{Second Wafer Results}%
\label{exp.fab2}



The next Al-STJ fabrication followed the same procedure as before but included a backside deep reactive ion etch (DRIE) to suspend the pixels on thin 0.5~\( \mu \)m\ SiN membranes.
A picture of a 128-pixel array of (130~\( \mu \)m)\(^2\) STJs from this wafer is shown in Figure~\ref{fig:fabricated-chip}, which is back-lit to highlight the transparent membranes.

\begin{figure}[!tb]
  \centering
  \vspace{-1em}
  \includegraphics[width=\linewidth]{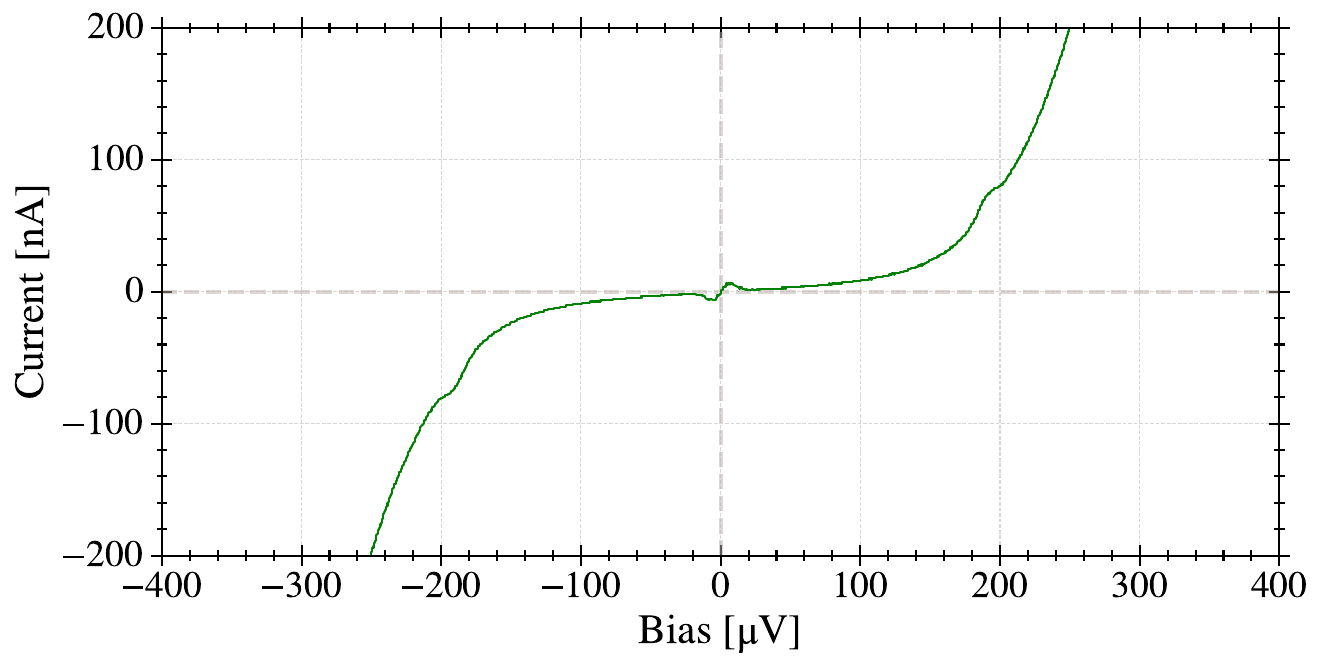}
  \caption{%
    Current-voltage characteristics of an Al-STJ on an SiN membrane at 100 mK. No thermal leakage current was visible, but an exponential current increase at \( 1\Delta_\text{Al}/e \) indicates some flux trapping in this run.%
    \label{fig:iv-fab2}
  }
  \vspace{-1em}
\end{figure}

The devices had good \( I(V) \) characteristics, with 10~nA leakage current and a dynamic resistance of 10~k\( \Omega \) at the bias point of 100~\( \mu \)V (Fig.~\ref{fig:iv-fab2}).
This is comparable to earlier Ta-STJs and sufficient for low-noise operation.
However, the device yield was low, with many pixels being shorted. We suspect that this is due to the high-voltage chuck that holds the wafer in place during backside etching. High-voltage discharge through a junction destroys the thin tunnel barrier, although not all STJs are affected. This suggests that the basic Al-STJ fabrication process on SiN membranes is sound as long as discharge risk is addressed during wafer mounting and backside etching.

Mitigating the \( \gamma \)-induced background is important for future phases of the BeEST, as this background can limit sensitivity to weaker signals as well as the total allowable activity across the chip.
While suspending STJs on membranes would greatly reduce this effect, another method was developed following the testing of these Al-STJs.
During Phase-III, analysis of the time-resolved events revealed that the \( \gamma \)-induced background could be completely removed by vetoing on events that were coincident in \(>\)4 pixels~\cite{BeEST-KimBray2025-PRD}.
As a result, optimizing STJ fabrication on SiN membranes became less of a priority.
However, it may be important to revisit this technique for later stages of the BeEST experiment with higher implant doses.

\subsection{Third Wafer Testing and Calibration}%
\label{exp.fab3}


The goal of the third wafer fabrication was to improve the energy resolution of the Al-STJs. The poor resolution of the first wafer was caused by a low responsivity that increased the influence of electronic noise.
Since the tunneling rate and thus the responsivity scales with the electrode thicknesses~\cite{DeKorVanDe1992-SPIE}, low responsivity may be addressed by making the electrodes thinner. However, this reduction is constrained by the requirement to absorb all \( ^7 \)Be atoms in the top Al electrode during implantation, which requires a minimum acceleration voltage between 5 and 10~kV.
Layer thicknesses of 200~nm for the Al base electrode and 100~nm for the Al absorber electrode were chosen as a safe compromise.
In addition, higher barrier transmissivity can further improve tunneling rates and responsivity. We have therefore reduced the oxygen exposure to increase the tunnel barrier transmissivity by a factor of \( \sim \)4. No back-side etching was performed on this wafer.




\begin{figure}[!bht]
  \centering
  \vspace{-1em}
  \includegraphics[width=\linewidth]{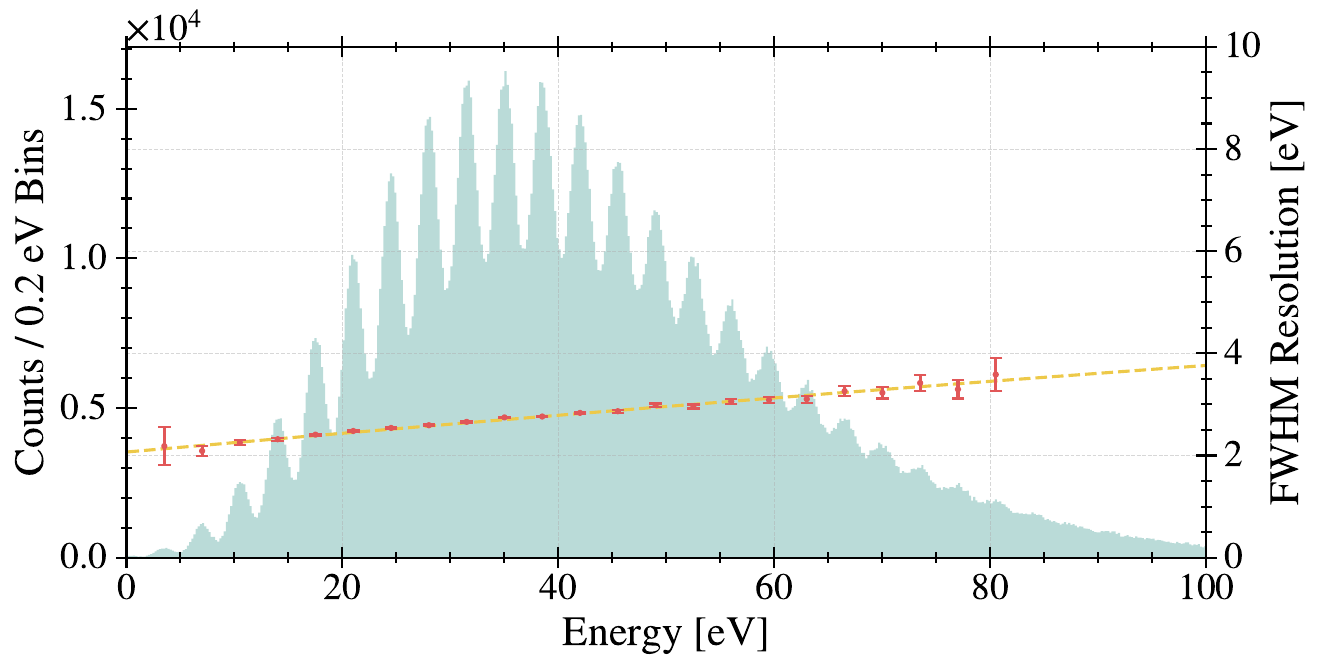}
  \caption{%
    Laser spectrum of the Al-STJ with increased responsivity.
    The resolution of 2--4~eV FWHM in the energy range of interest provides a measure of the intrinsic resolution, which can be used to quantify additional broadening sources in the \( ^7 \)Be recoil spectra of the BeEST experiment.%
    \label{fig:fab3-laser}
  }
  \vspace{-1em}
\end{figure}
The leakage current for these devices was around 50~nA with a dynamic resistance of 10~k\( \Omega \) at the bias voltage of 100~\( \mu \)V. When exposed to the 355~nm calibration laser at 100~Hz, these Al-STJs were able to resolve the individual peaks in the laser spectrum (Figure~\ref{fig:fab3-laser}). The devices demonstrated a responsivity of \( \sim \)100~nA/keV, resulting in a significant improvement in the signal-to-noise ratio over the first Al-STJs.
The resolved peaks were fit between 0--80~eV to a superposition of 25 Gaussian functions to extract centroids and peak widths, and the centroids were fit to a quadratic function to establish an energy scale. The energy resolution of this Al-STJ was between 2 to 4 eV below 100~eV. This is sufficient to calibrate the nuclear recoil spectrum in the BeEST experiment in the energy range of interest. It also provides a measure of the intrinsic linewidth of the detectors so that additional materials-dependent broadening components may be investigated.
%
%
%
\section{Summary}%
\label{discussion}

\begin{table*}[!htbp]
  \centering
  \caption{Comparison of Three Al-STJ Fabrication Runs}%
  \label{tab:fab_comparison}
  \begin{tabular}{r c c c c}
                                                                              &
    \multicolumn{0}{c}{\textbf{Parameter}}
                                                                              & \textbf{Fab 1}                 & \textbf{Fab 2}               & \textbf{Fab 3}                                                     \\
    \toprule
    \multirow{2}{*}{\textbf{Electrodes}}                                      & Base Electrode [nm, Al]        & 265                          & 265                           & 200 \parbox[c][11pt]{1cm}{}        \\
                                                                              & Top Electrode [nm, Al]         & 200                          & 200                           & 100                                \\
    \midrule
    \multirow{3}{*}{\parbox{1.5cm}{\raggedleft~\textbf{Fabrication Process}}} & Backside Etching               & No                           & Yes (DRIE, 0.5~\(\mu \)m SiN) & No                                 \\
                                                                              & Tunnel Barrier                 & Standard                     & Standard                      & 4\(\times \) higher transmissivity \\
                                                                              & Device Yield                   & High                         & Low (shorted pixels)          & High                               \\
    \midrule
    \textbf{Testing}                                                          & Laser Calibration              & No (Insufficient resolution) & Not attempted                 & Yes (0--80~eV)                     \\
    \midrule
    \multirow{2}{*}{\textbf{Noise}}                                           & Leakage Current [nA]           & 40--100                      & \(<\)10                       & \(\sim \)50                        \\
                                                                              & Dynamic Resistance             & Very high (\(\sim\infty \))  & 10~k\(\Omega \)               & 10~k\(\Omega \)                    \\
    \midrule
    \multirow{3}{*}{\textbf{Signal}}                                          & Responsivity [nA/keV\(^{-1}\)] & \(\sim \)20                  & Not measured                  & \(\sim \)100                       \\
                                                                              & Energy Resolution              & 9.2~eV FWHM                  & Not measured                  & 2.96~eV FWHM                       \\
                                                                              &                                & (K-GS at 108.5~eV)           &                               & (Laser at 50~eV)                   \\
    \midrule
    \multirow{2}{*}{\textbf{Goal}}                                            &                                & Proof of concept,            & SiN membrane                  & High resolution,                   \\
                                                                              &                                & nuclear spectra              & process demo                  & laser calibration                  \\
    \bottomrule
  \end{tabular}
\end{table*}

%
%

This work demonstrates the successful development of aluminum-based superconducting tunnel junction sensors for nuclear recoil spectroscopy.
We first showed that Al-STJs can be fabricated on SiN films.
These devices were used to measure the recoils from the decay of implanted \( ^7 \)Be.
Despite an electronic noise of \( \sim \)8~eV FWHM, the spectra provide initial evidence for material-dependent nuclear decay signatures that differ from Ta-STJs.
We then showed that, with minor developments on the fabrication process flow, we can backside-etch the Si wafer to produce high-quality Al-STJs on thin SiN membranes.
These membranes may be useful for future experiments where the total activity on a single chip results in an unmanageable rate of \( \gamma \)-induced background events.
Finally, we increased the responsivity of the Al-STJs with thinner electrodes and thinner tunnel barriers to reduce the influence of electronic noise.
These devices have achieved an energy resolution to 3.5~eV laser photons of 2--4 eV below 100~eV.

The resolution of the current iteration of Al-STJs is sufficient for nuclear recoil spectroscopy in the context of the BeEST experiment.
This enables future study of the material-dependent effects within the BeEST spectrum.
The characterization of material-dependent effects by systematic detector studies will be important to reduce systematic uncertainties in the search for sterile sub-MeV neutrino during the next phase of the BeEST experiment.
It will help distinguish material effects from a possible BSM physics signal in future experiments with improved statistics.

\section*{Acknowledgments}

The BeEST experiment is funded by the Gordon and Betty Moore Foundation grant 10.37807/GBMF11571, the DOE-SC Office of Nuclear Physics under awards DE-SC0021245 and DE-FG02--93ER40789, and the LLNL LDRD program grant 20-LW-006.
This research was performed under appointment to the Nuclear Nonproliferation International Safeguards Fellowship Program sponsored by the National Nuclear Security Administration's Office of International Nuclear Safeguards (NA-241).
This work was partially funded by Fundação para a Ciência e Tecnologia (FCT, Portugal) through national funds [UI/BD/151321/2021] and [LA/P/0117/2020].
TRIUMF receives federal funding via a contribution agreement with the National Research Council of Canada.
This work was performed under the auspices of the U.S. Department of Energy by LLNL under Contract DE-AC5207NA27344.
Francisco Ponce is funded as part of the Open Call Initiative at Pacific Northwest National Laboratory and conducted under the Laboratory Directed Research and Development Program.
Pacific Northwest National Laboratory is a multiprogram national laboratory operated by Battelle for the U.S. Department of Energy.

%

\bibliographystyle{IEEEtran}
\bibliography{Library-BibTeX}

\end{document}